\documentclass[12pt,showpacs,preprintnumbers,superscriptaddress,amsmath,amssymb,nofootinbib]{revtex4}
\usepackage{graphicx}% Include figure files
\usepackage{dcolumn}% Align table columns on decimal point
\usepackage{bm}% bold math
\usepackage{amssymb}
\usepackage{amsmath}
\usepackage{epsfig}    
\usepackage{color}
\usepackage{slashed}
\usepackage{hhline}
\def\be{\begin{equation}}
\def\ee{\end{equation}}
\newcommand{\bea}{\begin{eqnarray}}
\newcommand{\eea}{\end{eqnarray}}
\newcommand{\nn}{\nonumber}

\numberwithin{equation}{section}

\begin{document}

%%%%%%%%%
\title{Two loop Induced Dirac Neutrino Model and Dark Matters with Global $U(1)'$ Symmetry}
\preprint{KIAS-P14024}
\author{Hiroshi Okada}
\email{hokada@kias.re.kr}
\affiliation{School of Physics, KIAS, Seoul 130-722, Korea}

\begin{abstract}
We propose a two loop induced Dirac type neutrino model at TeV scale.
Subsequently, three types of dark matter particles; fermion and two bosons, are naturally introduced.
Here we discuss to analyze two possibilities; two component dark matter scenario (Dirac fermion and complex boson) and single dark matter one (another real boson), comparing to current experimental data such as Planck/WMAP and LUX.
We briefly mention the possibility to explain the discrepancy of the effective number of neutrino species reported by several experiments. 
%As for the two component scenario, we find a parameter set that is consistent with the observed relic density and current direct detection experiments.
%As for the single scenario, we find that the observed discrepancy of the effective number of neutrino species can be explained without any conflicts of the other phenomenologies in our scenario.

\end{abstract}
\maketitle
\newpage

\section{Introduction}
%%%Neutrinos and dark matters are expected to be an extension of the standard model (SM).
The nature of neutrinos is not yet known apparently  on even whether Majorana type or Dirac type in spite of several experiments of lepton flavor violating  processes such as neutrinoless double beta decay~\cite{Barabash:2011fn, Agostini:2013mzu,Auger:2012ar,Gando:2012zm,Alessandria:2011rc} as well as their own masses and their mixings~\cite{Tortola:2012te, Beutler:2014yhv}.
%%%
Furthermore, the nature of dark matter (DM) is not also known well in spite of many experiments such as direct detection searches({\it e.g.}, XENON100~\cite{Aprile:2012nq} and LUX~\cite{Akerib:2013tjd}), indirect detection searches({\it e.g.},  AMS-02~\cite{ams-02}, PAMELA~\cite{Adriani:2008zr, Adriani:2008zq}, Fermi-LAT~\cite{Bringmann:2012vr, Weniger:2012tx}, and XMN-Newton X-ray observatory~\cite{Bulbul:2014sua, Boyarsky:2014jta}), IceCube~\cite{Aartsen:2012kia}, and collider searches such as LHC~\cite{Giardino:2012dp}. These two issues could be the important tasks to be clarified in the future.

In a view of theoretical aspect, on the other hand,
radiatively induced neutrino models are one of the elegant solutions to implement a DM candidate within TeV scale~\cite{Ma:2006km, Aoki:2013gzs, Dasgupta:2013cwa, Krauss:2002px,
Aoki:2008av, Schmidt:2012yg, 
Bouchand:2012dx, Aoki:2011he, Farzan:2012sa,
Bonnet:2012kz, Kumericki:2012bf, Kumericki:2012bh, Ma:2012if, Gil:2012ya,
Okada:2012np, Hehn:2012kz, Dev:2012sg, Kajiyama:2012xg, Okada:2012sp,
Aoki:2010ib, Kanemura:2011vm, Lindner:2011it, Kanemura:2011mw,
Kanemura:2012rj, Gu:2007ug, Gu:2008zf, Gustafsson, Kajiyama:2013zla, Kajiyama:2013rla,
Hernandez:2013dta, Hernandez:2013hea, McDonald:2013hsa, Okada:2013iba,
Baek:2013fsa, Ma:2014cfa, Baek:2014awa, Ahriche:2014xra},~\cite{Ahn:2012cg, Ma:2012ez, Kajiyama:2013lja, Kajiyama:2013sza,
Ma:2013mga, Ma:2014eka}.
To achieve such kind of models, an additional local or global symmetry is always required to stabilize the DM candidate.
Notice here that the local symmetry requires a continuous symmetry, but the global one allows both a continuos and a discrete one.
Once one selects the global (continuous) symmetry, we should take in account  a phenomenology of the goldstone boson (GB), which could sometimes provide a promising explanation of the discrepancy of the effective 
number of neutrino species $\Delta N_{\rm eff}$~\cite{Weinberg:2013kea}.
When one chooses the local one, on the other hand, an exotic neutral gauged boson comes into our world, which gives us another phenomenological interests~\cite{Amaldi:1987fu, Mohapatra:2006gs}.
%If we expect that its symmetry should be a remnant one as a result of the symmetry breaking, a continuous symmetry like $U(1)$ group could be natural and promising.
In this paper, we show that introducing of a global continuous symmetry, which forbids some tree-level Yukawa Lagrangians, can be achieved in the framework of  two loop induced Dirac type neutrino\footnote{For another types of Dirac type neutrino, see, {\it i.e.}, Refs.~\cite{Davidson:2009ha, Luo:2008yc, Gu:2007gy,   Cheng:1980qt},~\cite{Aranda:2013gga, Memenga:2013vc},~\cite{Baek:2014qwa}.}.
After the global symmetry breaking spontaneously, a remnant symmetry can naturally identify the DM candidate for some mediated particles in neutrino masses. 
 %%%
% (Here we will consider to discuss a Dirac type neutrino model~\cite{Davidson:2009ha, Luo:2008yc, Gu:2007gy,   Cheng:1980qt},~\cite{Aranda:2013gga, Memenga:2013vc},~\cite{Baek:2014qwa}.) 
%%%

%In this letter, we propose a two-loop induced Dirac neutrino model to introducing some neutral particles with a global $U(1)$ charges. Then we will discuss DM property among our model.
This paper is organized as follows.
In Sec.~II, we show our model building including neutrino mass.
In Sec.~III, we analyze DM nature. We summarize and conclude in Sec.~VI.
%In appendices, we show the explicit Higgs potential and ...

%%%%%%%%%%%%%%%%%%%%%%%%%%%%%%%%%%%%%%
\section{The Model}
\subsection{Model setup}

\begin{table}[thbp]
\centering {\fontsize{10}{12}
\begin{tabular}{|c||c|c|c|c|c|}
\hline Particle & $L_L$ & $ e_{R} $  & $S_L$  &  $S_R$  & $N_R$ 
  \\\hhline{|=#=|=|=|=|=|$}
$(SU(2)_L,U(1)_Y)$ & $(\bm{2},-1/2)$ & $(\bm{1},-1)$  & $(\bm{1},0)$ & $(\bm{1},0)$ & $(\bm{1},0)$
\\\hline
%$U(1)_L$ & $+1$ & $-1$  & $-1$ & $+1$ & $-1$ & $0$ & $0$  & $0$  \\\hline
%%%
%$U(1)'$ & $-1/2$ & $-1/2$ & $-1/2$ & $-1/2$ & $3/2$  \\\hline
$U(1)'$ & $-2S-N$ & $-2S-N$ & $S$ & $S$ & $N$  \\\hline
%$Z_4\times Z_2$ & $(-1,-)$ & $(-1,-)$ & $(-i,-)$ & $(i,+)$ & $(-i,+)$ & $(-i,+)$ & $(i,+)$  & $(1,-)$  \\\hline
%%%
\end{tabular}%
} \caption{The particle contents and the charges for fermions.} 
\label{tab:1}
\end{table}

\begin{table}[thbp]
\centering {\fontsize{10}{12}
\begin{tabular}{|c||c|c|c|c|c|}
\hline Particle   & $\eta$  & $\Phi$  & $\chi_1$   & $\chi_2$  & $\Sigma$ 
  \\\hhline{|=#=|=|=|=|=|}
$(SU(2)_L,U(1)_Y)$ & $(\bm{2},1/2)$  & $(\bm{2},1/2)$   & $(\bm{1},0)$    & $(\bm{1},0)$     & $(\bm{1},0)$ \\\hline
%$U(1)_L$ & $0$ & $0$ & $0$  & $0$  \\\hline
%%%
%$U(1)' $  & $0$ & $0$ & $1$  & $-1$   & $2$  \\\hline
$U(1)' $  & $3S+N$ & $0$ & $-2S$  & $-S-N$   & $2(S+N)$  \\\hline
%$Z_4\times Z_2$  & $(i,-)$ & $(i,+)$ & $(i,-)$  & $(-i,+)$  \\\hline
%%%
\end{tabular}%
} \caption{The particle contents and the charges for bosons. }
\label{tab:2}
\end{table}

We discuss a two-loop induced radiative neutrino model. 
The particle contents are shown in Tab.~\ref{tab:1} and Tab.~\ref{tab:2}. 
We add three $SU(2)_L$ singlet vector like neutral fermions $S_L$ and
$S_R$, three singlet
Majorana fermions $N_R$.
For new bosons, we introduce $SU(2)_L$ doublet scalar $\eta$ and singlet scalars
$\chi_1$, $\chi_2$, and $\Sigma$ to the standard model (SM). 
We assume that  only the SM-like Higgs $\Phi$ and $\Sigma$ have vacuum
expectation values (VEVs). 
%Otherwise the $\mathbb{Z}_2$ symmetry which guarantees DM stability is spontaneously broken. 
The global $U(1)'$ symmetry is imposed so as to restrict their interaction
adequately and guarantee DM stability. Moreover, even after the $U(1)'$ symmetry is broken by the VEV of $\Sigma$, a
remnant symmetry of $Z_2$ retains which assures the stability.

The quantum number in the tables  can be driven as follows.
Let us at first define the $U(1)'$ charge $N$ for $N_R$ and $S$ for $S_{L/R}$.
Then one finds that all the terms are written in terms of those two charges.
%However, the charge of $\eta$, which is written $3S+N$, should be zero, in order not to couple Goldstone boson (GB) to Z-boson that is ruled out by the electroweak precision test.  As a result, $S$ and$N$ is fix to be $-1/2$ and $3/2$, respectively.
But several remarks are as follows:
\[S+N\neq0\  {\rm to\ forbid\ the\ term\ of}\ \bar L_L\Phi^\dag N_R, \]
\[S\neq0\  {\rm to\ forbid\ the\ term\ of}\  \Sigma^* \bar N^c_R N_R,\ \Sigma \chi_1(\chi_2)^2, \]
\[S+2N\neq0\  {\rm to\ forbid\ the\ term\ of}\  \Sigma \bar N^c_R N_R, \]
\[S\neq N\  {\rm to\ forbid\ the\ term\ of}\  \bar  L_L\eta^\dag N^c_R~\footnote{ $\bar  L_L\eta^\dag N^c_R$ does not affect to our model, but we assume to be zero for simplicity.}, \]
\[3S+2N\neq 0\  {\rm to\ forbid\ the\ term\ of}\  \Sigma^* \chi_1(\chi_2)^2~\footnote{If this term exists, we can generate a Majorana mass term for $N_R$ at two-loop level. As a result, neutrino mass is generated at four-loop level.
Such a model could include a little different feature from this model~\cite{preparation}.}. \]
Notice here that our charge assignment does not conflict with these conditions.
The five dimensional term $\Sigma^*\bar L_L \Phi^\dag N_R$ cannot be forbidden by any symmetries. However once the cut-off scale is taken to be GUT scale $\Lambda_{\rm GUT}\sim{\cal O}(10^{16})$ GeV, its Dirac mass scale is
$\langle\Phi\rangle \langle\Sigma\rangle/\Lambda_{\rm GUT}\le {\cal O}(0.01)$ eV (where we fix $\langle\Phi\rangle=246$ GeV and $\langle\Sigma\rangle=$1000 GeV), which can be naturally tiny than the active neutrino mass ${\cal O}(0.1-1)$ eV.
We impose $S+N=1$ for our convenience, as we will discuss later.

The renormalizable Lagrangian for Yukawa sector and scalar potential under these assignments
are given by
\begin{eqnarray}
\mathcal{L}_{Y}
&=&
y_\ell \bar L_L \Phi e_R +
y_{\eta} \bar L_L \eta^\dag S_R +
y_{\chi_1}  \bar S^c_L S_L \chi_1 +
y_{\chi_2}  \bar S^c_R N_R \chi_2 +M_{S}\bar S_L S_R+\rm{h.c.} \\ 
%%%
\mathcal{V}
&=& 
 m_1^{2} \Phi^\dagger \Phi + m_2^{2} \eta^\dagger \eta  + m_3^{2}
 \Sigma^\dagger \Sigma  + m_4^{2} \chi^\dagger_1 \chi_1  + m_5^{2} \chi^\dagger_2 \chi_2\nn\\
 &&
+ \mu [\Sigma(\chi_2)^2+{\rm h.c.}]
 + \lambda_0[(\Phi^\dag \eta)(\chi_1\chi_2)+{\rm h.c.}]
 \nn\\
&&
  +\lambda_1 (\Phi^\dagger \Phi)^{2} + \lambda_2 
(\eta^\dagger \eta)^{2} + \lambda_3 (\Phi^\dagger \Phi)(\eta^\dagger \eta) 
+
 \lambda_4 (\Phi^\dagger \eta)(\eta^\dagger \Phi)
%+\lambda_5 [(\Phi^\dagger \eta)^{2} + \mathrm{h.c.}]+
%%%
%\lambda'_5 [(\Sigma^\dagger \chi)^{2} + \mathrm{h.c.}]+\lambda''_5 [(\Sigma \chi)^{2} + \mathrm{h.c.}]
\nn\\&&+
\lambda_6 (\Sigma^\dagger \Sigma)^{2} 
+\sum_{i=1,2}\lambda'^{(i)}_6 (\chi^\dagger_i \chi_i)^{2} 
+\sum_{i=1,2}\lambda''^{(i)}_6 \left(\Sigma^\dagger\Sigma\right)\left(\chi^\dag_i\chi_i\right) 
\\&&+
\lambda_7  (\Sigma^\dagger \Sigma)(\Phi^\dagger \Phi ) 
+ \sum_{i=1,2}\lambda'^{(i)}_7(\chi^\dagger_i \chi_i)(\Phi^\dagger \Phi) +
 \lambda_8  (\Sigma^\dagger \Sigma) (\eta^\dagger \eta)
 + \sum_{i=1,2}\lambda'^{(i)}_8
(\chi^\dagger_i \chi_i) (\eta^\dagger \eta) ,\nn
%+\left[a(\eta^\dag \Phi)(\Sigma\chi)+{\rm  h.c.}\right]+\left[a'(\Phi^\dag \eta)(\Sigma\chi)+{\rm h.c.}\right],
\label{HP}
\end{eqnarray}
where the first term of $\mathcal{L}_{Y}$ can generates the charged-lepton masses, and $\mu$ and $\lambda_0$ can be chosen to be real
without any loss of 
generality by renormalizing the phases to scalar bosons. The couplings
$\lambda_1$, $\lambda_2$, $\lambda_6$ and 
$\lambda'^{(i)}_6$ have to be positive
to stabilize the Higgs potential.  
Inserting the tadpole conditions; $m^2_1=-\lambda_1v^2-\lambda_7v'^2/2$ and
 $m^2_3=-\lambda_6v'^2 - \lambda_7v^2/2$,
the resulting mass matrix of the neutral component of $\Phi$ and 
$\Sigma$ defined as 
\begin{equation}
%\mathrm{with}\quad
\Phi^0=\frac{v+\phi^0(x)}{\sqrt{2}},\qquad
\Sigma=\frac{v'+\sigma(x)}{\sqrt{2}}e^{iG(x)/v'},\quad
%\mathrm{after~the~symmetry~breaking.}
\end{equation}
 is given by
\begin{equation}
m^{2} (\phi^{0},\sigma) = \left(%
\begin{array}{cc}
  2\lambda_1v^2 & \lambda_7vv' \\
  \lambda_7vv' & 2\lambda_6v'^2 \\
\end{array}%
\right) \!=\! \left(\begin{array}{cc} \cos\alpha & \sin\alpha \\ -\sin\alpha & \cos\alpha \end{array}\right)
\left(\begin{array}{cc} m^2_{h} & 0 \\ 0 & m^2_{H}  \end{array}\right)
\left(\begin{array}{cc} \cos\alpha & -\sin\alpha \\ \sin\alpha &
      \cos\alpha \end{array}\right), 
\end{equation}
where $v=$246 GeV, $h$ implies SM-like Higgs and $H$ is an additional CP-even Higgs mass
eigenstate. The mixing angle $\alpha$ is given by 
\be
\tan 2\alpha=\frac{\lambda_7 v v'}{\lambda_6 v'^2-\lambda_1 v^2}.
\ee
The Higgs bosons $\phi^0$ and $\sigma$ are rewritten in terms of the mass eigenstates $h$ and $H$ as
\begin{eqnarray}
\phi^0 &=& h\cos\alpha + H\sin\alpha, \nn\\
\sigma &=&- h\sin\alpha + H\cos\alpha.
\label{eq:mass_weak}
\end{eqnarray}
A goldstone boson $G$ appears due to the spontaneous symmetry breaking of
the global $U(1)'$ symmetry. 
%This massless particle would be dark radiation contributing to the effective neutrino number we will discuss later. 

Each mass eigenstate for the inert Higgses is given as
\begin{eqnarray}
m_\eta^2\equiv m^{2} (\eta^{\pm}) &=& m_2^{2} + \frac12 \lambda_3 v^{2}
 + \frac12 \lambda_8 v'^{2}, \\ 
m^2_{\eta^{0}} &=& m_2^{2} + \frac12
 \lambda_8 v'^{2} 
 + \frac12 (\lambda_3 + \lambda_4) v^{2}, \\ 
 %%%
m^2_{\chi_{1}} &=& m_4^{2} 
+ \frac12\left( \lambda''^{(1)}_6 v'^2+  \lambda'^{(1)}_7 v^2\right), \\
%%%
m^2_{\chi_{2R}} &=& m_5^{2} 
+ \frac12\left( \lambda''^{(2)}_6 v'^2+  \lambda'^{(2)}_7 v^2 +2\sqrt2\mu v'\right), \\  
m^2_{\chi_{2I}} &=& m_5^{2} 
+ \frac12\left( \lambda''^{(2)}_6 v'^2+  \lambda'^{(2)}_7 v^2 -2\sqrt2\mu v'\right).
\end{eqnarray}
Notice here that $\eta^0$ and $\chi_1$ are complex scalar neutral bosons.

\subsection{Neutrino mass matrix}

%%%%%%%%%%%%%%%%%%%
\begin{figure}[cbt]
\begin{center}
%\unitlength=1mm
%\hspace{-2cm}
%\begin{picture}(40,100)
%\includegraphics[width=6cm]{neutrino.eps}
\includegraphics[scale=0.6]{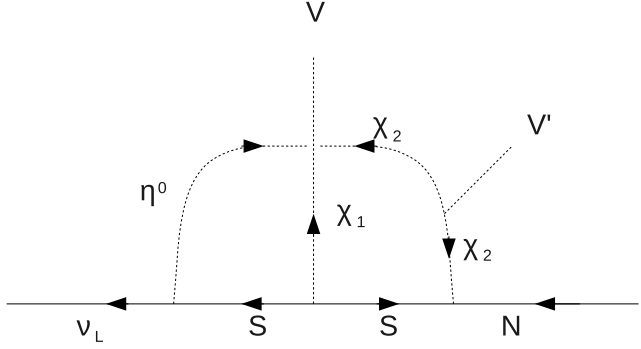}
%\qquad
% \includegraphics[scale=1]{neutrino2.eps}
%   \end{picture}
   \caption{Radiative generation of neutrino masses.  }
   \label{neutrino-diag}
\end{center}
\end{figure}
%%%%%%%%%%%%%%%%%%%
The Dirac neutrino mass matrix at two-loop level as depicted in
the left hand side of Fig.~\ref{neutrino-diag} is given by 
\begin{equation}
(m_{\nu})_{ab}=
\frac{\lambda_0v}{4}\left[  \frac{(y_\eta)_{ac}(M_S)_{c}(y_{\chi_1})^*_{cd}(M_S)_d(y_{\chi_2})_{db}}{2\sqrt2(4\pi)^4[(M_{S})^2_{c}-m^2_{\eta^0}]} \right] 
\left[F(x_{ia})|_{\chi_{2R}}
-F(x_{ia})|_{\chi_{2I}}\right],
\label{eq:neutrinomass}
%- (R\to I),
\end{equation}
where the loop function $F$ is computed by
\begin{equation}
F(x_{ia})
=\int_0^1dy \int_0^{1-y}dz
%\\\nn&&
\frac{-x_{3d}\ln(x_{3c})+(\alpha-x_{3d})\left(\frac{1-y-z-x_{2c}y-x_{c1}z}{y-(1-z)^2-x_{1c}z}\right)
\ln\left(\frac{\alpha(z-z^z)} {1-y-z-x_{2d}y-x_{1d}z}\right)
}
{1-y-z+x_{2d} y+\{x_{1d}+x_{3d}(z-1)\}z},
\end{equation}
where $\alpha\equiv(M_{Sc}/M_{Sd})^2$, $x_{ia}\equiv(m_{\chi_i}/M_{Sa})^2$ with $\chi_3\equiv \eta^0$ and the indices of $x$ are defined as $i=(1,2,3)$ and $a=(c,d)$~\footnote{One can find the original Zee--Babu type neutrino formula in the limit of $M_S\to0$ and $x_3\to x_2$~\cite{AristizabalSierra:2006gb}.}.
%%%
One finds rather wide allowed range to explain the neutrino masses reported by Planck data ~\cite{Ade:2013lta}; $m_{\nu}<0.933~\mathrm{eV}$, with the following parameters:  $(y_{\eta}y^*_{\chi_1}y_{\chi_2})={\cal O}$(0.1), $M_{S}={\cal O}$(500) GeV, $m_{\chi_1}={\cal O}(500)$ GeV, $m_{\eta^0}={\cal O}(1000)$ GeV, $\mu={\cal O}$(0.1) GeV, and $v'={\cal O}$(1000) GeV, and $\lambda_0={\cal O}$(0.5).

%%%
%From the neutrino mass formula, $(y_{\eta}y^*_{\chi_1}y_{\chi_2})\approx 10^{-4}$ is needed to obtain the proper neutrino mass scale by assuming $M_S\approx 0.5~\mathrm{TeV}$ and $\mathcal{O}(0.1)$ of the loop functions. 

%%%%%%%%%%%%%%%%%%%
%\begin{figure}[tbc]\begin{center}\includegraphics[scale=0.8]{neumass-plot.eps}
%   \caption{Order estimation for the sum of the neutrino masses as a function of the $\chi_{2I}$ mass. Here we fix $(y_{\eta}y^*_{\chi_1}y_{\chi_2})\approx 0.5$, $M_{S}=$500 GeV, $m_{\chi_1}=$500 GeV, $m_{\eta^0}=$2500 GeV, $\mu=$0.1 GeV, and $v'=$1000 GeV. The black line represents $\lambda_0=$1.0, the green line represents $\lambda_0=$0.5, and the blue line represents $\lambda_0=$0.1. The red line; $\sum m_\nu=$0.933, is the upper bound reported by the experiment of Planck~\cite{Ade:2013lta}.  }   \label{neut-plot}\end{center}\end{figure}
%%%%%%%%%%%%%%%%%%%

{\it Lepton Flavor Violations (LFVs)}:
$\mu\to e\gamma$ process gives the most stringent bound. 
The upper limit of the branching ratio is given by
$\mathrm{Br}\left(\mu\to e\gamma\right)\leq5.7\times 10^{-13}$ at 95\%
confidence level from the MEG experiment~\cite{meg2}.

Our contribution to the $\mu\to e\gamma$ process only comes from the coupling of $y_\eta$ and its branching ratio can be computed as
\begin{equation}
\mathrm{Br}\left(\mu\to e\gamma\right)=
\frac{3\alpha_{\mathrm{em}}}{64\pi G_F^2m_\eta^4}
\left|\sum_{i}\left(y_\eta\right)_{i\mu}
\left(y_{\eta}\right)_{ie}^*F_2\left(\frac{M_{Si}^2}{m_\eta^2}\right)\right|^2 ,
\label{eq:lfv}
\end{equation}
where $\alpha_{\mathrm{em}}=$1/137 is the fine structure constant, $G_F$
is the Fermi constant and $F_2(x)$ is the loop function defined in
ref.~\cite{Ma:2001mr}. 

Ad can be seen in these  Eq.~(\ref{eq:neutrinomass}) and Eq.(\ref{eq:lfv}), we can avoid this constraint very easily by taking that $y_\eta$ is diagonal. This is because neutrino sector has a lot of free parameters such as $y_{\chi_1}$ or $y_{\chi_2}$ , from which we could obtain observed mixings as well as active neutrino masses~\cite{Tortola:2012te}.

%\newpage

\section{Dark Matter}
\label{sec:DM}
We have three DM candidates: the lightest one of three vector like fermions $S$, the lightest
one of $\chi_1$(complex scalar) and $\chi_2$(real scalar), as a result of the remnant symmetry $Z_2$ after the breaking of $U(1)'$ symmetry. Here we consider 
$S$ and $\chi_1$ as multicomponent DM scenario, since they do not decay into SM particles at lading order.
Also we consider $\chi_{2I}$ as a single DM scenario\footnote{
Since the property of  $\chi_{2R}$ and $\chi_{2I}$ is the same, we focus on the $\chi_{2I}$ as a DM candidate taking positive sign of $\mu$.}. %The sign of $\mu$ determines which one of  $\chi_{2R/I}$ is DM
Notice here that neutral $\eta$ component is ruled out by the direct detection through $Z$-boson particle, since it is a complex scalar. Here we discuss to analyze two cases: multicomponent DMs scenario ($S, \chi_1$)  and single DM scenario $\chi_2$.

\subsection {Multicomponent Dark Matter scenario}
At first, we will discuss the relic density of DMs; $\Omega h^2\approx$0.12, reported by Planck~\cite{Ade:2013lta}.
%%%
The DM ($S$) can annihilate into the other DM ($\chi_1$), but cannot
decay into the SM particles with the renormalizable interactions. 
%They cannot be taken care independently when one computes each relic density since one DM annihilates into the other DM. 
 We have to compute the set of Boltzmann equations in order to obtain the correct relic density of those two DMs. 
The set of Boltzmann equations is written as 
\begin{eqnarray}
\frac{dn_S}{dt}+3Hn_S&=&
-\langle\sigma_{S}{v_{\rm rel}}\rangle\left(n_S^2-{n_S^{\mathrm{eq}}}^2\right)
+\langle\sigma_{\mathrm{ex}}{v_{\rm rel}}\rangle\left[n^2_{\chi_1}-\left(\frac{n_\chi^{\mathrm{eq}}}
{n_S^{\mathrm{eq}}}\right)^2n_S^2\right],\\
\frac{dn_{\chi_1}}{dt}+3Hn_{\chi_1}&=&
-\langle\sigma_{\chi_1}{v_{\rm rel}}\rangle\left(n^2_{\chi_{1}}-{n_{\chi_1}^{\mathrm{eq}}}^2\right)
-\langle\sigma_\mathrm{ex}{v_{\rm rel}}\rangle\left[n^2_{\chi_1}-\left(\frac{n^{\mathrm{eq}}_{\chi_1}}
{n_S^{\mathrm{eq}}}\right)^2 n_S^2\right],
\end{eqnarray}
where the time of universe is expressed by $t$, $n_S$ and $n_{\chi_1}$ are the
number density of $S$ and $\chi_1$ respectively. 
The thermally averaged annihilation cross section into all channels is written as
$\langle\sigma_{S}{v_{\rm rel}}\rangle$ for $S$. 
For $\chi_1$, the total cross section into the SM particles is written  
by $\langle\sigma_{\chi}{v_{\rm rel}}\rangle$.
 $\langle\sigma_{\mathrm{ex}}{v_{\rm rel}}\rangle$
is the cross section of the DM exchange process $\bar SS\to \chi^*_1\chi_1$. Notice here that we assume $m_{\chi_1}\le2M_S$, otherwise $\chi_1$ can decay into $2S$. 
%If $\langle\sigma_{\mathrm{ex}}{v_{\rm rel}}\rangle$ is negligible compared with
%$\langle\sigma_{\chi_1}{v_{\rm rel}}\rangle$, the simultaneous Boltzmann equation becomes independent of each other, and the total relic density should be a sum of $N_1$ and $\chi_R$ : $\Omega_{N_1}h^2+\Omega_{\chi_R}h^2$. 

{\it Fermionic DM( $S$)}:
%%%%%%%%%%%%%%%%%%%
%\begin{figure}[tbc]\begin{center}
%\includegraphics[scale=0.8]{S-relic.eps}
 %  \caption{ Relic density as a function of the DM mass, where we fix $v'$=1000 GeV, $m_{\eta}=$2500 GeV and $y_\eta=$1.}   \label{s-relic}\end{center}\end{figure}
%%%%%%%%%%%%%%%%%%%
%$S$ can contribute to the observed relic density; $\Omega h^2\approx$0.12, reported by Planck~\cite{Ade:2013lta}.
The dominant cross section for $S$ is obtained through %$t$- and $u$-channel of $S$ in GB pairs and
$t$-channel of $\eta$ in the limit of massless final state lepton pairs as follows:
\bea
%(\sigma v_{\rm rel})(\bar S S\to 2G)&\approx&\frac{5 M^2_S}{6144\pi v'^4}v_{\rm rel}^2,\\
%%%
(\sigma v_{\rm rel})(\bar S S\to \bar\ell\ell)&\approx&
\frac{|y^\dag_{\eta}y_{\eta}|^2}{128\pi M_{S}^2(1+x_3)^2}
\left[1-\frac{1-x^2_3+3 x_3}{3(1+x_3)^2}v_{\rm rel}^2\right],
\eea
where $M_{\rm DM}$ is the mass of $S$, $1\le x_3=m^2_{\eta}/M^2_S$, and we assume $m_{\eta}=m_{\eta^0}$ for simplicity.
%We find that there  exists a wide range of allowed region as can be seen in Fig.~\ref{s-relic}.

{\it  Bosonic DM ($\chi_{1}$) }:
%$\chi_1$ can contributes to the relic density and direct detection through Higgses.For the  relic density,
There are four final state annihilation modes at tree level: $\chi_1\chi^*_1\to
hh,\: ZZ, W^+W^-,\: f\overline{f}$. Each cross section is
given in Ref.~\cite{Baek:2013fsa}.

{\it Exchange contribution of $\bar S S\to\chi^*_1\chi_1$}: 
%%%
The DM exchange channel $\bar S S\to\chi^*_1\chi_1$ via t-channel is found as
\begin{eqnarray}
\sigma_{\mathrm{ex}}{v}_{\mathrm{rel}}\left(\bar S S\to\chi^*_1\chi_1\right)
&\approx&
  \frac{|y_{\chi_1}|^4}{128\pi (m^2_{\chi_{1}}-2M^2_S)^2}
%%%
\sqrt{1-\frac{m^2_{\chi_1}}{M^2_S}}\times \nn\\
 %%%
&& \left[- m^2_{\chi_{1}} + M^2_S
 -
\frac{m^6_{\chi_{1}} -6 m^4_{\chi_{1}} M^2_S+20 m^2_{\chi_{1}}M^4_S}{24 (m^2_{\chi_{1}}-2M^2_S)^2}v^2_{\rm rel}
 %%%
 \right]
\label{eq:tansit},
\end{eqnarray}

\begin{figure}[tbc]\begin{center}
\includegraphics[scale=0.8]{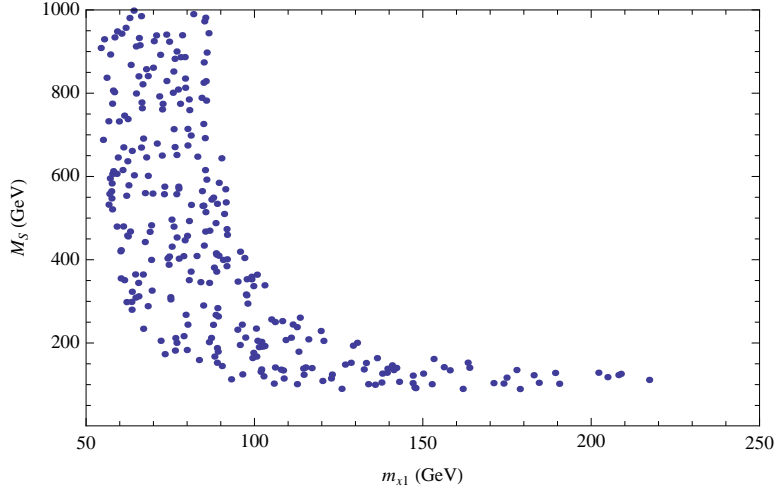}
 \caption{Allowed regions of DM masses to obtain the observed Relic density $\Omega h^2\approx$0.12, where $m_{\chi_1}\le2M_S$ is assumed to forbid the rapid decay between DMs.}   \label{multi-relic}\end{center}\end{figure}

{\it Parameter set as a solution of the relic density}:
We simply show an allowed region  to obtain a observed  relic density $\Omega h^2\approx$0.12
under the following cross sections  in Fig.~\ref{multi-relic}:
\begin{eqnarray}
&&
%m_{\chi_1}=M_S=500\ {\rm GeV},\quad
 a_{S}=4\times 10^{-9}\ {\rm GeV}^{-2},\quad 
 b_{S}=2.5\times 10^{-12}\ {\rm GeV}^{-2},
\quad
 a_{\chi_1}=4\times 10^{-9}\ {\rm GeV}^{-2},
  \nn\\
 &&
    b_{\chi_1}=4\times 10^{-12}\ {\rm GeV}^{-2},\quad 
  a_{\rm ex}=4\times 10^{-10}\ {\rm GeV}^{-2},  
 \quad  b_{\rm ex}=0, 
  \label{relic-para}
\end{eqnarray}
where each of $a_i$ and $b_i$ is the the s-wave contribution and the p-wave one ($i=S,\chi_1,{\rm ex}$).
As can be seen from Fig.~\ref{multi-relic}, we obtain
\be
50\ {\rm GeV}\lesssim M_S\lesssim1000\ {\rm GeV},\quad50\ {\rm GeV}\lesssim m_{\chi_1}\lesssim250\ {\rm GeV}.
\ee

%%%%%%%%%%%%%%%%%%%
%%%%%%%%%%%%%%%%%%%

%%%%%%
{\it Direct detection}: 
Only $\chi_1$ DM candidate can contribute to
the spin independent elastic cross section  that can be obtained through neutral Higgses as
\begin{equation}
\sigma_p=\frac{C\mu_{\chi}^2m_p^2}{\pi m_{\chi_1}^2 v^2}
\left(\frac{\mu_{\chi\chi h}\cos\alpha}{m_h^2}
+\frac{\mu_{\chi\chi H}\sin\alpha}{m_H^2}\right)^2,
\end{equation}
where $\mu_\chi$ is reduced mass defined as
$\mu_\chi=(m_{\chi_1}^{-1}+m_p^{-1})^{-1}$, $m_p=938~\mathrm{MeV}$ is the proton
mass and $C\approx0.079$. The elastic cross section is constrained by LUX as $\sigma_p\lesssim
{\cal O}(10^{-45})~\mathrm{cm^2}$ 
at around the point $m_{\chi_1}\approx{\cal O}(100)~\mathrm{GeV}$~\cite{Akerib:2013tjd}. 
%%%
The cubic
couplings $\mu_{\chi\chi h}$ and $\mu_{\chi\chi H}$ are the three point vertex of $\chi_{1}\chi_{1}h$ and $\chi_{1}\chi_{1}H$ with a mass dimension, and can be written as a function of  ($\lambda_1, \lambda_6, \lambda'^{(1)}_6,\lambda''^{(1)}_6, \lambda_7, \lambda'^{(1)}_7,v,v', \alpha$), which are not proportional to the term of $\mu$. Hence it is easy to satisfy the constraint of the direct detection experiments, by controlling $\mu_{\chi\chi h}$ and $\mu_{\chi\chi H}$.
%The couplings $\mu_{\chi\chi h}$ and $\mu_{\chi\chi H}$ are given by 
%\begin{eqnarray}
%\mu_{\chi\chi h}&=&-\left(\lambda_5'+\lambda_5''+\frac{\lambda_6''}{2}\right)v'\sin\alpha
%+\frac{\lambda_7'}{2}v\cos\alpha, \\\mu_{\chi\chi H}&=&\left(\lambda_5'+\lambda_5''+\frac{\lambda_6''}{2}\right)v'\cos\alpha+\frac{\lambda_7'}{2}v\sin\alpha. \end{eqnarray}
%Notice here that the constrained cross section will increased(=be relaxed), since the number density of $\chi_1$ dominates half the component of the whole relic density as can be seen in our benchmark point in Eq.~(\ref{relic-para}).

\subsection{Bosonic DM $\chi_{2I}$ }
{\it Relic density}: The dominant contribution for the relic density comes from the GB boson final state through the  four-point interaction, $s$-channel, and $t(u)$-channel due to the $\mu$ term
\footnote{$\chi_{2R}$, of course, has the same annihilation channels as the $\chi_1$.}.
Its thermal averaged cross section can be then obtained as 
%\bea
\be
(\sigma v_{\rm rel})\approx
\frac{M_{\rm DM}^2 (m_{\chi_{2R}}^4 +4M_{\rm DM}^2-5m_{\chi_{2R}}^2M_{\rm DM}^2-4\sqrt{2}\mu v' m_{\chi_{2R}}^2-4\sqrt{2}\mu v' M_{\rm DM}^2)^2 }{64\pi v'^4(m^2_{\chi_{2R}}+ M_{\rm DM}^2)^2(m^2_{\chi_{2R}}-4 M_{\rm DM}^2)^2},
\ee
%\eea
where  $M_{\rm DM}$ is the mass of $\chi_{2I}$, we abbreviate the $p$-wave due to the complicated form.
%, and take a rather concrete $U(1)'$ charge assignment  $S+N=1$ for our convenient~\cite{preparation}.

%%%%%%%%%%%%%%%%%%%
%\begin{figure}[tbc]
%\begin{center}
%\includegraphics[scale=0.8]{chi2-relic.eps}
   %\caption{ Relic density as a function of the DM mass, where we fix $\mu=$0.1 GeV, and $v'=$1000 GeV.}
%   \label{chi2-relic}\end{center}\end{figure}
%%%%%%%%%%%%%%%%%%%

%%%%%%
{\it Direct detection}: 
The spin independent elastic cross section can be obtained through neutral Higgses as
\begin{equation}
\sigma_p=\frac{C\mu_{\chi}^2m_p^2}{\pi M_{\rm DM}^2 v^2}
\left(\frac{\mu_{\chi_{2I}\chi_{2I} h}\cos\alpha}{m_h^2}
+\frac{\mu_{\chi_{2I}\chi_{2I} H}\sin\alpha}{m_H^2}\right)^2,
\end{equation}
where $\mu_\chi$ is reduced mass defined as
$\mu_\chi=(M_{\rm DM}+m_p^{-1})^{-1}$, $m_p=938~\mathrm{MeV}$ is the proton
mass and $C\approx0.079$. Here each of $\mu_{\chi_{2I}\chi_{2I}h}$ and $\mu_{\chi_{2I}\chi_{2I} H}$ is the three point vertex of $\chi_{2I}\chi_{2I}h$ and $\chi_{2I}\chi_{2I}H$ with a mass dimension, and can be written as a function of  ($\mu, \lambda'^{(2)}_6,\lambda''^{(2)}_6, \lambda'^{(2)}_7,v,v', \alpha$).
%The couplings $\mu_{\chi\chi h}$ and $\mu_{\chi\chi H}$ are given by 
%\begin{eqnarray}
%\mu_{\chi\chi h}&=&-\left(\lambda_5'+\lambda_5''+\frac{\lambda_6''}{2}\right)v'\sin\alpha
%+\frac{\lambda_7'}{2}v\cos\alpha, \\\mu_{\chi\chi H}&=&\left(\lambda_5'+\lambda_5''+\frac{\lambda_6''}{2}\right)v'\cos\alpha+\frac{\lambda_7'}{2}v\sin\alpha. \end{eqnarray}
%The elastic cross section is constrained by LUX as $\sigma_p\lesssim 7.6\times10^{-46}~\mathrm{cm^2}$ at the most stringent point $M_{\rm DM}\approx33~\mathrm{GeV}$~\cite{Akerib:2013tjd}. 
One finds that there exists wide allowed region to satisfy the observed relic density and the constraint of the direct detection experiments due to the similar property of , using the same bench parameter set as those of neutrino sector. %as can be seen in Fig.~\ref{chi2-relic}, too.

Thus these quartic couplings are required to be $\mathcal{O}(0.5)$ in order to satisfy
the constraint when $v'\sim1~\mathrm{TeV}$ and $\sin\alpha\sim1$. 
Due to the strong constraint from direct detection of DM, 
the annihilation cross section for the process $\chi_{2I}\chi_{2I}\to
f\overline{f}$ via Higgs s-channel~\footnote{Notice here that $2Z$ or $W^\pm$ final state mode does not appear in the limit of $\alpha=0$.} is extremely
suppressed, which is given by 
\begin{equation}
\sigma{v}_{\mathrm{rel}}=\frac{y_f^2}{2\pi}\left(1-\frac{4m_f^2}{s}\right)^{3/2}
\left|\frac{\mu_{\chi_{2I} \chi_{2I}  h}\cos\alpha}{s-m_h^2+im_h\Gamma_h}
+\frac{\mu_{\chi_{2I} \chi_{2I}  H}\sin\alpha}{s-m_H^2+im_H\Gamma_H}\right|^2,
\end{equation}
where $s\approx 4m_{\chi_{2I}}^2(1+v_{\mathrm{rel}}^2/4)$, $\Gamma_h$ and
$\Gamma_H$ are the decay width of $h$ and $H$. 
This is because GB final sate is the dominant.

%\textcolor{red}{
{\it $\Delta N_{\rm eff}$ }:
The discrepancy of the effective 
number of neutrino species $\Delta N_{\rm eff}$ has been reported by several
experiments such as Planck~\cite{Ade:2013lta}, WMAP9
polarization~\cite{Bennett:2012zja}, and ground-based
data~\cite{Das:2013zf, Reichardt:2011yv}, 
which tell us $\Delta N_{\rm eff}=0.36 \pm 0.34$ at the 68~\% confidence level. 
%%%
Such a deviation $\Delta N_{\rm eff}\approx0.39$ is achieved due to GB in our model, if
the following condition can be satisfied~\cite{Baek:2013fsa}:
\be
\frac{\sin^22\alpha(m^2_h-m^2_H)^2m^7_\mu m_{\rm pl}}{4(vv')^2(m_h m_H)^4}\approx1,
\ee
where where $m_{\rm pl} \approx 1.2 \times 10^{19}$ GeV is the Planck mass and $m_\mu\approx$105.7 MeV is the muon mass. 
It implies that an extra neutral boson $H$ to be tiny ${\cal O}$(500) MeV, and $\alpha$ is small enough.
As a result, the DM mass should be less than ${\cal O}$(5) GeV.
%%%
This could be achieved by our scenario in a different parameter set~\cite{Baek:2013fsa},
 since the dominant relic density of our DM does not include such a light extra Higgs.
%Hence we can explain the observed  $\Delta N_{\rm eff}$ by taking quartic couplings related to direct detection to be tiny enough. 
%%%}

\section{Conclusions}
We have constructed a two-loop induced Dirac neutrino model with a global $U(1)'$ symmetry, in which we have naturally
introduced DMs; Dirac fermion and neutral  scalar bosons. Due to several Yukawa couplings related to neutrinos, we can easily control such parameters to avoid any LFV processes like a $\mu\to e,\gamma$. 
%%%

We have analyzed two possibilities of the DM candidate; two component scenario with $S$ and $\chi_1$, and single boson one $\chi_{2I}$.
As for two component scenario, we have computed the Boltzmann equation explicitly 
depicted the figure of the observed relic density in terms of two DM masses with a fixed parameter set of the cross section in Fig.~\ref{multi-relic}.
%and shown an parameter set to obtain the observed relic density.
We have also discussed the direct detection, in which it is easy to satisfy the current bound due to some free parameters that are not related to the relic density.
%%%
As for bosonic DM ($\chi_{2I}$), 
we have shown that there exists a solution to satisfy the observed relic density and the direct detection, since some parameters that are used to each main channel are separate.
%we have discussed that the elastic spin independent scattering cross section can be  within a current direct detection searches such as LUX that provides a most stringent upper bound, as well as its relic density. 

%%
Also we have brief mentioned the possibility to explain the observed discrepancy of the effective 
number of neutrino species.
%, in whcih without any conflicts of the other phenomenologies that we have discussed in our previous letter.

%\fi

%\newpage
%%%%%%%%%%%%%%%%%%%%%%%%%%%%%%%%%%%
\vspace{0.5cm}
%\hspace{0.2cm} {\bf Acknowledgments}
\section*{Acknowledgments}
\vspace{0.5cm}
Author thanks to Prof. Seungwon Baek and Dr. Takashi Toma for fruitful discussions.
%%%%%%%%%%%%%%%%%%%%%%%%%%%%%%%%%%%

\end{document}